# Finite-size scaling in two-dimensional superfluids

Norbert Schultka and Efstratios Manousakis

*Department of Physics and Center for Materials Research and Technology*
*Florida State University, Tallahassee, Florida 32306*

October 14, 1993

**Abstract**

Using the $x - y$ model and a non-local updating scheme called cluster Monte Carlo, we calculate the superfluid density of a two dimensional superfluid on large-size square lattices $L \times L$ up to $400 \times 400$. This technique allows us to approach temperatures close to the critical point, and by studying a wide range of $L$ values and applying finite-size scaling theory we are able to extract the critical properties of the system. We calculate the superfluid density and from that we extract the renormalization group beta function. We derive finite-size scaling expressions using the Kosterlitz-Thouless-Nelson Renormalization Group equations and show that they are in very good agreement with our numerical results. This allows us to extrapolate our results to the infinite-size limit. We also find that the universal discontinuity of the superfluid density at the critical temperature is in very good agreement with the Kosterlitz-Thouless-Nelson calculation and experiments.

## 1 Introduction

The singular behavior in the thermodynamic functions of liquid $^4He$ close to the superfluid transition can be understood in terms of a complex order parameter $\psi(\vec{r})$ which is the ensemble average of the helium atom boson creation operator. This ensemble average is defined inside a volume of size much greater than the interatomic distance but much smaller than the temperature-dependent coherence length. In order to describe the physics at longer length scales, which is important very close to the critical point, we need to consider spatial fluctuations of the order parameter. These fluctuations can be taken into account by assigning a Landau-Ginzburg free energy functional $\mathcal{H}(\psi(\vec{r}))$ to each configuration of $\psi(\vec{r})$ and performing the sum of $e^{-\mathcal{H}/k_B T}$ over such configurations. The power laws governing the long distance behavior of the correlation functions and the critical exponents associated with the singular behavior of the thermodynamic quantities close to the critical point are insensitive to the precise functional form of $\mathcal{H}[\psi]$, and they are the same for an entire class of such functionals. The following Landau-Ginzburg free-energy functional can be used to describe the fluctuations of the complex order parameter $\psi$

$$\mathcal{H}[\psi(\vec{r})] = \int d^d x \big(\frac{1}{2}|\nabla \psi|^2 + \frac{1}{2}m_0|\psi|^2 + \frac{1}{4}\lambda_0|\psi|^4\big). \tag{1}$$

Another form of the Landau-Ginzburg free energy is the planar $x - y$ model which is expressed as

$$\mathcal{H} = J \sum_{\langle i,j \rangle} \vec{s}_i \cdot \vec{s}_j, \tag{2}$$

where the summation is over all nearest neighbors, $\vec{s} = (\cos\theta, \sin\theta)$ is a two-component vector which is constrained to be on the unit circle. The angle $\theta$ corresponds to the phase of the order parameter $\psi(\vec{r})$. It can be shown that the model (2) and the Landau-Ginzburg free-energy (1) belong to the same universality class.

The idea of a spatially varying order parameter is crucial in order to understand the critical properties of helium films. For example, in two dimensions, macroscopic order in the sense of a nonzero average of $\psi(\vec{r})$ is eliminated by thermal fluctuations[1]. Namely the destruction of the order is due to phase fluctuations of the local order parameter. In a film, we are interested in two-dimensional(2D) transitions occurring at temperatures below the corresponding three-dimensional(3D) lambda critical temperature $T_\lambda$, where amplitude fluctuations around the Ginzburg-Landau minimum are small. However, a constant phase change costs no free energy, because of Goldstone modes. Such long-wavelength fluctuations have very small free energy and in two dimensions they destroy the long-range order. On the other hand, configurations which correspond to the well-known vortices are responsible for the phase transition in thin films of liquid $^4He$. These configurations play the key role in the Kosterlitz-Thouless[2] theory where the transition can be understood in terms of unbinding of quantized vortices of opposite sign.

The two-dimensional $x-y$ model has been studied both analytically and numerically (see for example[2]-[10]). First of all, Kosterlitz and Thouless (KT) included the contribution of vortex excitations by mapping the model to a two-dimensional gas of interacting vortices and by using an approximate renormalization-group theory. Due to the non-perturbative nature of the topological KT phase transition it is difficult to develop an analytical method which allows us to calculate the corrections to the KT calculation. Numerical simulation studies (see for example [8]) seem to indicate that the KT theory is both qualitatively and to a good degree of approximation quantitatively accurate. However, early Monte Carlo studies were hindered by the so-called critical slowing down where the autocorrelation time for local-updating schemes grows very rapidly with the system size as one approaches the critical region. Thus, with local-updating Monte Carlo methods one can only study small size systems close to the critical region. More recently[11] a non-local updating scheme, the so-called cluster Monte Carlo, has been proposed which very effectively deals with the problem of critical slowing down. This method has been used to test the Kosterlitz-Thouless scenario by calculating the correlation length in the 2D $x-y$ model[9].

In this paper we carry out a detailed finite-size scaling analysis of the superfluid density $\rho_s(T, L)$ in pure 2D helium films of size $L \times L$ with respect to $L$ using the cluster Monte Carlo updating technique. The superfluid density is directly accessible to experiments[12] and is characterized by interesting scaling behavior with film thickness[13]. There are some earlier studies[6, 7, 14] of the superfluid density (helicity modulus) in the $x-y$ model using the local Metropolis Monte Carlo method. In this paper we shall perform a thorough study of this quantity on large enough size lattices, and by deriving finite-size scaling forms we are able to extrapolate to the infinite size lattice. In order to calculate the superfluid density in large finite-size films close enough to the critical point we use the cluster Monte Carlo method[9, 11]. We first calculate the renormalization group beta-function from the finite-size scaling of the superfluid density. The calculated beta-function is then used to collapse our calculated $\rho_s(T, L)$ for all $L$ on one universal curve. Our results obey finite-size scaling using values for the critical exponents close to those calculated by Kosterlitz and Thouless. We have used the KT theory and the renormalization group equations for the superfluid density and chemical potential obtained by Nelson and Kosterlitz[3] to derive the dependence of the superfluid density on $L$ below and at the critical temperature. Our results obey faithfully these finite-size scaling laws and we use them to extrapolate to the infinite size lattice at all temperature values used to calculate the superfluid density. We obtain an accurate value for the ratio $\rho_s(T_c)/T_c$ of the discontinuity of the superfluid density at the critical point which within our error bars is in very good agreement with the value obtained by Nelson and Kosterlitz and with the experimental results[12].

The $x-y$ model can also be used to describe certain spin-systems where the superfluid density corresponds to the spin stiffness. In fact, all our results obtained in this paper can find applications in describing the critical properties of such spin-systems or any system whose critical properties can be described by a two component order parameter and can be mapped to the $x-y$ model.

In the next section we discuss the calculation of the superfluid density $\rho_s(L,T)$ using the cluster Monte Carlo method. In section III, we discuss the finite-size scaling above the critical temperature $T_c$ and the calculation of the renormalization group (RG) beta function. In section IV, we use the RG equations to derive a finite-size dependence of $\rho_s(L,T)$ and we use that to extrapolate our numerical results to the infinite-size lattice. In the last section we draw some conclusions and discuss future extensions of this work.

## 2 Formulation and Monte Carlo Calculation

Within the formalism of the $x-y$ model the physical quantity that corresponds to the superfluid density is the helicity modulus $\Upsilon(T)$. The helicity modulus was introduced by Fisher, Barber and Jasnow in order to define the coherence length in superfluid helium[15]. Let us consider liquid helium confined in a cylindrical domain of cross-sectional area $A$ and length $H$. We twist the order parameter in the upper boundary layer by a small angle $\phi$ with respect to the lower boundary. The helicity modulus measures the change in the free energy due to this twist and is defined as[15]

$$\Upsilon(T) = \lim_{A,H\to\infty} \frac{2H}{\phi^2 A}(F(T,\phi) - F(T,0)), \qquad (3)$$

where $F(T,\phi)$ and $F(T,0)$ denote the free energy of the system with the twist and without it, respectively. Since a superfluid flux is introduced by the twist, a connection to the superfluid density can be established [15]

$$\rho_s(T) = \left(\frac{m}{\hbar}\right)^2 \Upsilon(T), \qquad (4)$$

where $m$ is the mass of the helium molecule.

The definition (3) can be rewritten [17] as

$$\Upsilon(T) = \left.\frac{\partial^2 F(T,k)}{\partial k^2}\right|_{k=0}, \qquad (5)$$

where $k$ denotes the wave vector of the twist. The last definition can be easily applied to any dimension. By working in a rotating reference frame in the spin space[17] and using (5), one finds [14, 16]

$$\begin{aligned}\frac{\Upsilon_\mu}{J} &= \frac{1}{V}\left\langle \sum_{\langle i,j\rangle} \cos(\theta_i - \theta_j)(\vec{e}_\mu \cdot \vec{e}_{ij})^2 \right\rangle \\ &\quad - \frac{\beta}{V}\left\langle \left(\sum_{\langle i,j\rangle} \sin(\theta_i - \theta_j)\vec{e}_\mu \cdot \vec{e}_{ij}\right)^2 \right\rangle,\end{aligned} \qquad (6)$$

where $V$ is the volume of the lattice, $\vec{e}_\mu$ is the unit vector in the corresponding bond direction, and $\vec{e}_{ij}$ is the vector connecting the lattice sites $i$ and $j$. As we work only on $L\times L$ lattices, i.e. in an isotropic system, we will omit the vector notation for the helicity modulus in the following. The above thermal averages are calculated as follows:

$$\langle O\rangle = Z^{-1}\int \prod_i d\theta_i\, O[\theta]\exp(-\beta\mathcal{H}). \qquad (7)$$

$O[\theta]$ expresses the dependence of the observable $O$ on the configuration $\{\theta_i\}$ and the partition function $Z$ for the model is given by

$$Z = \int \prod_i d\theta_i \, \exp(-\beta \mathcal{H}), \tag{8}$$

where $\beta = 1/k_B T$. These expectation values were computed by means of a Monte Carlo simulation using Wolff's cluster algorithm[11], which effectively deals with the problem of critical slowing down[9]. We calculated the superfluid density on lattices of sizes $L \times L$ with periodic boundary conditions (BC) where $L = 20, 30, 40, 60, 100, 300, 400$. For a given temperature we performed of the order of $10^4$ steps to reach thermalization and of the order of $10^6$ measurements. Our calculations were carried out on a heterogeneous environment of workstations which include Sun, IBM RS/6000 and DEC alpha AXP workstations and on the Cray-YMP supercomputer and took several months of CPU time.

Our results for the helicity modulus as a function of temperature T (in units of $J$) for various size-lattices are summarized in Fig.1. Notice that the finite-size effects are very strong. It is clear from this figure that in order to obtain an accurate value for $T_c$ and an accurate value for the universal discontinuity of $\Upsilon(T_c)/T_c$, we need a careful finite-size scaling analysis. The dashed line is our extrapolated values for $\Upsilon(L \to \infty, T)$ and the approach leading to that will be discussed in the next two sections. The finite-size dependence of $\Upsilon(L,T)$ is different above and below $T_c$. Above $T_c$, the infinite square lattice value of $\Upsilon$ is zero; however, the correlation length grows in a very singular way as known from the Kosterlitz-Thouless theory:

$$\xi(T) \propto \exp[B(T - T_c)^{-1/2}], \tag{9}$$

thus, finite-size effects become important for $T < T^*$ where

$$T^* - T_c \sim (\frac{B}{\ln(L)})^2 \tag{10}$$

which explains the very slow approach of $\Upsilon(L, T)$ to its infinite-$L$ value above $T_c$. Below $T_c$, the finite-size effects on $\Upsilon$ are weaker, but the value of the universal discontinuity strongly depends on the value of $T_c$. In the next section we discuss the finite-size scaling above $T_c$. In section IV, we discuss the finite-size effects below $T_c$ and how to extrapolate our results to the infinite-$L$ limit and obtain a value for $T_c$.

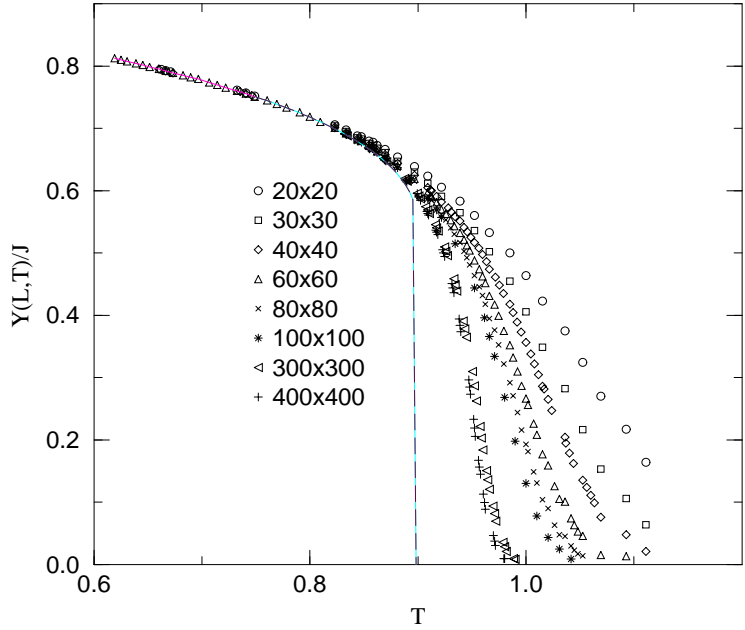

Figure 1: The helicity modulus $\Upsilon(L,T)$ as a function of $T$ for various lattices $L \times L$. The error bars are ommited because they are smaller than the size of the symbols.

## 3  The beta function and finite-size scaling above $T_c$

The helicity modulus $\Upsilon(L, T)/J$ for a 2D $x - y$ model is dimensionless and, thus, should be kept constant under scaling transformations. Namely, the beta function can be obtained by defining a function $T = F(L)$ such that $\Upsilon(L', F(L')) = \Upsilon(L, F(L)) = \Upsilon_p$ where $\Upsilon_p$ is a constant, a physical value of the helicity modulus. The beta function is defined as:

$$\beta(T) = -\lim_{L \to \infty} \frac{dT}{d\ln(L)}. \tag{11}$$

For large enough $L$, where finite-size effects influencing the calculation of the beta-function are small, the Kosterlitz-Thouless theory suggests the following form of $\beta(T)$:

$$\begin{aligned} \beta(T > T_c) &= c(T - T_c)^{1+\nu} \\ \beta(T \leq T_c) &= 0. \end{aligned} \tag{12}$$

Inserting this ansatz for $\beta(T)$ into the expression (11) and integrating we obtain:

$$\ln L - \frac{B}{(T - T_c)^\nu} = z, \tag{13}$$

where $B = 1/(\nu c)$ and $z$ is a constant of integration depending on the value of $\Upsilon$ used to define the scaling transformation. Eq. (13) defines the scale transformation which leaves the physical observables invariant. If the helicity modulus is considered as a function of $z$, then all curves for various large enough lattice sizes should collapse on the same universal curve. This conclusion can also be reached in a different way: The various values of the dimensionless observable $\Upsilon/J$ for different lattices $L \times L$ as a function of the ratio $L/\xi(T)$, where $\xi(T)$ is the Kosterlitz-Thouless correlation length given by Eq. (9), should collapse onto the same universal curve. The same will happen if one plots the calculated values of $\Upsilon(L,T)/J$ as a function of the variable $z = \ln(L/\xi(T))$. Notice that the latter is identical to the expression (13).

The beta function can be determined as follows[8, 18, 19]. For a pair of lattices $L_1, L_2$ let us consider the calculated $\Upsilon(L_1, T)$ and $\Upsilon(L_2, T)$ for all values of $T$. Choosing a value $\Upsilon$ for the helicity modulus we can determine two pairs $(T_1, L_1)$ and $(T_2, L_2)$ such that $\Upsilon = \Upsilon(L_1, T_1) = \Upsilon(L_2, T_2)$. Using these two points we calculate a value of the beta-function at $\bar{T} = (T_1 + T_2)/2$ as follows

$$\beta(\bar{T}) = -\frac{T_2 - T_1}{\ln(L_2) - \ln(L_1)}. \tag{14}$$

By choosing a different value of the helicity modulus we can obtain a new value for $\beta(\bar{T})$ at a different value of $\bar{T}$. Fig.2 shows the beta function for the pair $L_1 = 300$ and $L_2 = 400$.

We then fit the calculated $\beta(T)$ to the form given by the Kosterlitz-Thouless expression (12) using $c, T_c$ and $\nu$ as fitting parameters. The result of the fit is $c = 1.10(52), \nu = 0.56(28)$ and $T_c = 0.883(16)$. If we now plot the two curves $\Upsilon(L_1, T)$ and $\Upsilon(L_2, T)$ as a function of the variable $z = \ln L - B/(T - T_c(L))^\nu$ (i.e, we plot $\Upsilon(L_1, T)$ versus $\ln L_1 - B/(T - T_c)^\nu$ and $\Upsilon(L_2, T)$ versus $\ln L_2 - B/(T - T_c)^\nu$ on the same plot) they should collapse on the same curve. This is demonstrated in Fig.3.

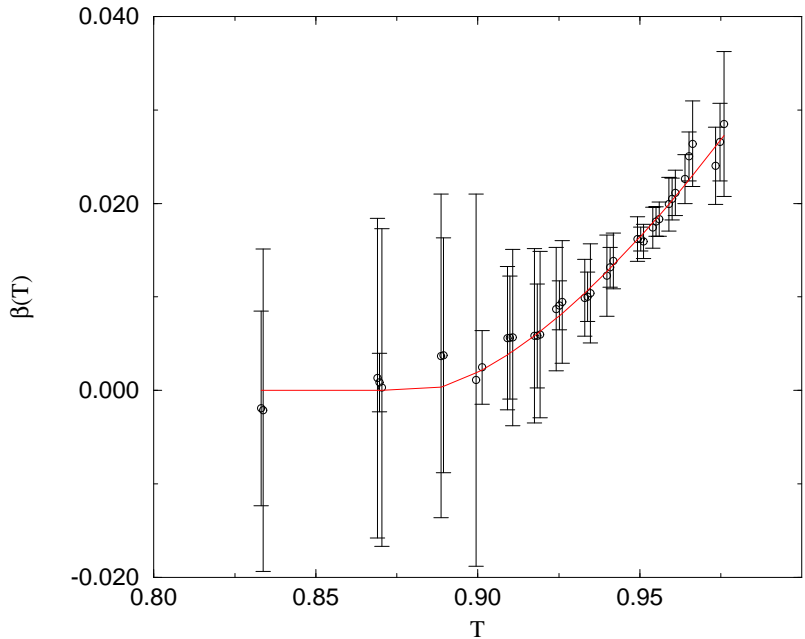

Figure 2: The beta function obtained from the $300 \times 300$ and $400 \times 400$ lattices.

We have repeated this procedure for all the pairs $(L_1, L_2) = (30, 40), (40, 60), (60, 80), (80, 100), (100, 300), (300, 400)$ and the results for the fitting parameters $c, T_c,$ and $\nu$ are given in Table 1.

| $L_1, L_2$ | $c$ | $\nu$ | $T_c$ |
|---|---|---|---|
| 30,40 | 0.906(93) | 0.535(82) | 0.8509(81) |
| 40,60 | 0.891(64) | 0.502(58) | 0.8621(49) |
| 60,80 | 1.12(25) | 0.75(21) | 0.846(16) |
| 80,100 | 1.15(36) | 0.56(23) | 0.875(16) |
| 100,300 | 1.00(15) | 0.550(99) | 0.8757(57) |
| 300,400 | 1.10(52) | 0.56(28) | 0.883(16) |

Table 1: Fitted values of the parameters(12) of the beta function (14) for each lattice pair.

For large enough values of $(L_1, L_2)$ the results of these different fits should be the same, and as can be seen from Table 1 the values of these parameters are approximately the same. However, a more careful observation tells us that the pseudocritical temperature $T_c$, obtained from different lattice pairs, has a slight monotonic increase with system size. In the short report of Ref.[20] we used all the values of the beta function obtained from all pairs of lattices with $L \leq 100$ to fit them to one beta function curve given by (12).

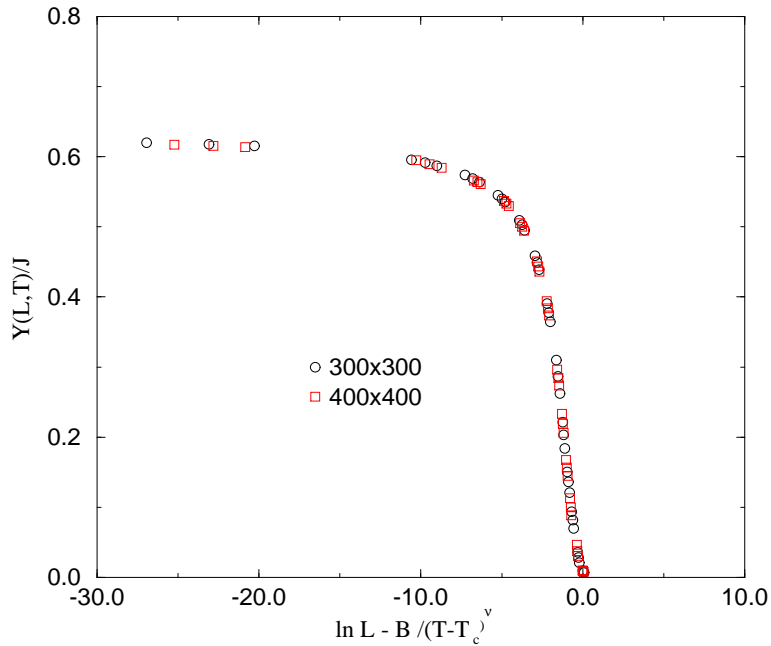

Figure 3: The helicity modulus $\Upsilon(L,T)$ as a function of $z$ for the $300 \times 300$ and the $400 \times 400$ lattices. The error bars are ommited because they are smaller than the size of the symbols.

This gives an average value of $T_c$ and ignores this slight but systematic increase of $T_c$ with size. If we ignore this dependence of $T_c$ on the size we can collapse our calculated values of $\Upsilon(L,T)$ for all size lattices by finding one set of parameters that fit all the data for $\beta(T)$ obtained from the lattices considered here. This, however, ignores the systematic dependence of the pseudocritical temperature on the size of the lattices used to extract the beta function and thus yields a poorer determination of the critical temperature as compared to the estimate of $T_c$ obtained by using the largest possible $L$. Thus, the conclusion of this study of the finite-size scaling of $\Upsilon(L,T)$ above $T_c$ is that the most accurate lower bound for $T_c$ is the one obtained for the largest size lattices used in this work and it is $T_c = 0.883(16)$. Next, however, we shall study the scaling of the superfluid density below $T_c$ and we shall provide a more accurate way to estimate $T_c$.

## 4  Finite-size scaling below $T_c$. The ratio $T/\Upsilon$

Let us consider the dimensionless ratio

$$K(L,T) = T/\Upsilon(L,T), \qquad (15)$$

In Table 2 and Table 3 the values of $K(L,T)$ for different temperatures below $T_c$ and various lattice sizes are given.

We wish to extrapolate our calculated values to the limit $L \to \infty$. For that we need an extrapolation formula and we shall derive one using the RG calculation of Nelson and Kosterlitz[3]. These RG equations

|   | $L$ | | | |
|---|---|---|---|---|
| $T$ | 20 | 30 | 40 | 60 |
| 0.8783 | 1.33549(61) | 1.34840(83) | 1.35610(84) | 1.3637(13) |
| 0.8772 | 1.33130(61) | 1.34374(62) | 1.35119(62) | 1.35852(63) |
| 0.8760 | 1.32693(80) | 1.3391(10) | 1.3463(14) | 1.3534(15) |
| 0.8696 | 1.30467(39) | 1.31553(40) | 1.32093(40) | 1.32697(20) |
| 0.8547 | 1.25691(37) | 1.26491(37) | 1.26886(38) | 1.27226(38) |
| 0.8333 | 1.19457(34) | 1.19939(17) | 1.20215(35) | 1.20407(17) |
| 0.7407 | 0.97839(13) | 0.97930(13) | 0.97982(13) | 0.980244(91) |
|  | $K(L,T)$ | | | |

Table 2: Calculated values for $K(L,T)$ for different temperatures and lattice sizes.

|   | $L$ | | | |
|---|---|---|---|---|
| $T$ | 80 | 100 | 300 | 400 |
| 0.8783 | 1.3677(17) | 1.3718(26) |  | 1.382(17) |
| 0.8772 | 1.36252(42) | 1.36634(42) |  | 1.37642(86) |
| 0.8760 | 1.3574(17) | 1.3609(27) |  | 1.371(15) |
| 0.8696 | 1.33104(41) | 1.33287(41) | 1.33924(41) | 1.34006(62) |
| 0.8547 | 1.27453(38) | 1.27567(38) | 1.27853(57) | 1.27949(77) |
| 0.8333 | 1.20530(14) | 1.20563(17) | 1.20808(35) | 1.20703(52) |
| 0.7407 | 0.980244(78) | 0.980373(78) |  |  |
|  | $K(L,T)$ | | | |

Table 3: Calculated values for $K(L,T)$ for different temperatures and lattice sizes

are

$$\frac{dK(l,T)}{dl} = 4\pi^3 y^2(l,T), \quad (16)$$

$$\frac{dy(l,T)}{dl} = (2 - \pi K^{-1}(l,T))y(l,T), \quad (17)$$

where $l = \ln L$, and $\ln y$ is the chemical potential for creating a single vortex. We wish to solve these RG equations for a finite length scale $L$ on a square lattice. The solution for finite-$L$ will correspond to a square of size $L \times L$ with free BC. The infinite-$L$ value of the superfluid density is clearly independent of the BC. In addition, we believe that different BC will only give a different value for the constant prefactor of the leading finite-$L$ correction, but the same exponent of the leading $L$-dependent correction.

To solve Eqs (16) and (17) we chose the initial conditions to be [3]

$$K(0,T) = T/\Upsilon_0,$$
$$y(0,T) = y_0 \exp(-\kappa/K(1,T)). \quad (18)$$

The precise values of the constants $\Upsilon_0, y_0$, and $\kappa$ are of no importance to us here. The solution to (16) and (17) can be written as

$$x - \ln x - 2\pi^2 y^2 = C, \quad (19)$$

where $x = 2K/\pi$ and $C$ is a constant depending on the initial conditions. The topologically "ordered" phase is given by the set of curves, for which $y \to 0$ but $x \leq 1$ [3]. The critical temperature is the largest temperature where this condition can be fulfilled. For the corresponding curve the constant $C = 1$ and $T_c$ is obtained via $x_c$, which itself solves

$$x_c - 1 - \ln x_c = 2\pi^2 \exp(-2/x_c). \tag{20}$$

We set $y_0 = 1$ and $\kappa = \pi/2$ (see also [4]).

Now, let us find the solution $x(l, T)$ close to the critical temperature $T_c$ in the limit $l \to \infty$. From (16) and (19) we obtain

$$\frac{dx}{dl} = 4(x - C - \ln x). \tag{21}$$

Introducing

$$x = \bar{x} + \tilde{x}, \tag{22}$$

with

$$\bar{x} - C - \ln \bar{x} = 0, \tag{23}$$

and expanding in $\tilde{x}$ up to the second order, we have

$$\tilde{x}(l, T < T_c) = \bar{x}\left(1 + \frac{2(1 - \bar{x})}{1 - \tilde{c}\exp(4l(1/\bar{x} - 1))}\right), \tag{24}$$

$$\tilde{x}(l, T_c) = -\frac{1/2}{l + \tilde{c}}. \tag{25}$$

Here $\tilde{c}$ is a constant of integration. In terms of $K_\infty(T) \equiv K(L \to \infty, T)$ and $L$ these equations take the form

$$K(L \to \infty, T) = K_\infty(T)\left(1 + \frac{2(1 - K_\infty(T)/K_c)}{1 - \tilde{c}L^{4(K_c/K_\infty(T) - 1)}}\right), \tag{26}$$

$$K(L, T_c) = K_c\left(1 - \frac{1/2}{\ln L + \tilde{c}'}\right). \tag{27}$$

Here $K_c = \pi/2$ for the Kosterlitz-Thouless-Nelson approach. The leading correction due to finite-$L$ below $T_c$ is given by:

$$K(L \to \infty, T) = K_\infty(T)\left(1 + \tilde{D}(1 - K_\infty(T)/K_c) L^{-4(K_c/K_\infty(T) - 1)}\right), \tag{28}$$

where the constant prefactor $\tilde{D}$ depends on the BC. Notice that the $L$-independent factor $(1 - K_\infty(T)/K_c)$ is needed in order to obtain the same extrapolated value at $T_c$ if one takes the infinite-$L$ limit first and the $T \to T_c$ limit afterwards or vice versa. For the values of $L$ and $T$ used in our calculation the two forms (26) and (28) are very close because $L^{4(K_c/K_\infty(T) - 1)} \gg 1$. Thus, in our calculation we have obtained the same values for the extrapolated $K_\infty(T)$ within error bars by using either form. In the tables we have chosen to give the results obtained with (26).

First using Eq. (26) with $K_c$ as a fitting parameter we can extrapolate to the $L \to \infty$ limit for a given $T$. Fig.4 shows a typical fit.

In Table 4 the results of our fits to the form (26) are presented.

We find an average value $\bar{K}_c = 1.558 \pm 0.059$. Since this value of $K_c$ is, within error bars, the same as the one obtained by Nelson and Kosterlitz, we fixed the value of $K_c = \pi/2$ and performed another set of extrapolations to the $L \to \infty$ using (26) with $\tilde{c}$ as the only fitting parameter. The results of this fit are given in Table 5 and the extrapolated values of $K(L \to \infty, T)$ are the same within error bars as those obtained by letting $K_c$ be a free parameter.

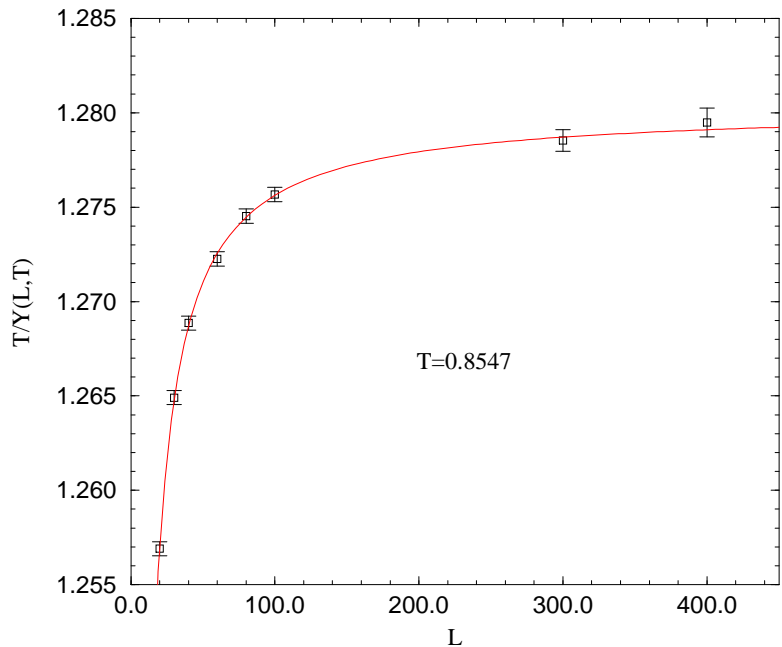

Figure 4: $T/\Upsilon(L,T)$ as a function of $L$ at $T = 0.8547$. The solid curve is the fit to (26).

After having determined the values for $K_\infty(T)$ for different temperature values, we used the form:

$$K_\infty(T \to T_c) = K'_c \left[1 - b\left(1 - \frac{T}{T_c}\right)^{1/2}\right], \qquad (29)$$

to fit our values for $K_\infty(T)$ in order to determine the discontinuity and the value of $T_c$. This form can be derived from the Nelson and Kosterlitz RG equations where the values for $K'_c = \pi/2$ and $b \sim 0.5$. Table 6 gives the results of three different fits.

The first fit includes all 7 data points for $K_\infty(T)$ given in Table 4. This fit is shown in Fig.5.

In the second fit we have excluded the point that corresponds to the lowest temperature. In the third fit we have excluded the two points that correspond to the lowest two temperature values. All three fits give within error bars the same values of $K'_c, b$, and $T_c$. Note, that $K'_c = \pi/2$ cannot be ruled out within error bars.

Since within our error bars $K'_c = K_c = \pi/2$ in agreement with the RG equations we repeated the procedure described above by fixing the value of these two parameters to $\pi/2$. The results of this fitting procedure are given in Table 7.

Here we find that the fit which includes the temperature $T = 0.7407$ is much worse (compare the values of $\chi^2$ in Table 7) than the other two fits. Therefore we believe that Eq. (29) is not a good approximation for $T \simeq 0.75$. Thus our fits give values of $K_c$ and $K'_c$ very close to $\pi/2$ and all our values of $T_c$ are given in the range $T_c = 0.895 \pm 0.004$ Thus, our values for the critical temperature are in good agreement with other critical temperature estimates (see for example [10], where $T_c = .894(5)$ is given).

| $T$ | $K(L=\infty, T)$ | $\tilde{c}$ | $K_c$ |
|---|---|---|---|
| 0.8783 | 1.3902(22) | 1.266(043) | 1.5984(78) |
| 0.8772 | 1.3849(9) | 1.309(32) | 1.5885(54) |
| 0.8760 | 1.3788(50) | 1.34(12) | 1.583(27) |
| 0.8696 | 1.3448(7) | 1.350(53) | 1.5800(78) |
| 0.8547 | 1.2803(3) | 1.211(40) | 1.5926(44) |
| 0.8333 | 1.2083(2) | 1.565(99) | 1.5300(82) |
| 0.7407 | 0.98045(5) | 1.115(75) | 1.4331(4) |

Table 4: Fit results to Eq. (26) using $K_c$ and $\tilde{c}$ as fitting parameters.

| $T$ | $K(L=\infty, T)$ | $\tilde{c}$ |
|---|---|---|
| 0.8783 | 1.3962(88) | 1.37(11) |
| 0.8772 | 1.3876(12) | 1.393(32) |
| 0.8760 | 1.381(37) | 1.38(73) |
| 0.8696 | 1.3456(6) | 1.410(47) |
| 0.8547 | 1.2810(3) | 1.384(27) |
| 0.8333 | 1.2077(1) | 1.156(26) |
| 0.7407 | 0.98033(5) | 0.263(18) |

Table 5: Fit results to Eq. (26) taking $K_c = \pi/2$.

$K(L, T)$ at $T_c$ for various values of $L$ obey reasonably well the form (27). Fitting $K(L, T_c)$ to this expression ((27)) yields $K_c \simeq \pi/2$ and $\tilde{c}' = -1.18 \pm 0.26$.

## 5 Summary

We have thoroughly investigated the finite-size scaling properties of the superfluid density of liquid helium in a pure two-dimensional geometry above and below the critical temperature. We found that the ratio $T/\Upsilon$ feels very strong finite-size effects. By solving the RG equations (16) and (17) for finite $L$ we were able to find the leading finite-size correction below and at $T_c$. Our numerically calculated values of $T/\Upsilon$ faithfully obey these finite-size scaling forms and thus we can safely extrapolate to the infinite-size lattice. The obtained values for the ratio $T/\Upsilon$ within error bars were found to be in very good agreement with the Kosterlitz-Thouless-Nelson theory and experimental results.

Having been able to keep the finite-size effects due to the finite $L$ under control, we now plan to study films of finite thickness, namely of size $L \times L \times H$ where $L >> H$. We need to remove the finite-size effects due to finite $L$ and then study the finite-size scaling of the superfluid density with $H$. From theoretical and experimental investigations (see for example [12],[13]) we are lead to the conclusion, that at a certain crossover temperature all helium films start behaving as two-dimensional systems. In a film geometry the dimensionless ratio $T/\Upsilon$ has to be replaced by the dimensionless ratio $T/(\Upsilon H)$. Since it is expected to exhibit finite-size effects with respect to the planar extensions of the film, it will be necessary to extrapolate to the infinite plane. The finite-size scaling should be governed by Eq. (26) and (27). The infinite planar size values of $T/(\Upsilon H)$ should behave according to Eq. (29). We are in the process of extending our work to such finite-thickness helium films and we shall discuss our finding in relationship to the experimental findings[12, 21].

| data points | $K_c'$ | $b$ | $T_c$ | $\chi^2$ |
|---|---|---|---|---|
| 7 | 1.6163(31) | 0.9390(33) | 0.8984(5) | 0.69(71) |
| 6 | 1.5673(82) | 0.890(13) | 0.8924(10) | 0.77(82) |
| 5 | 1.622(57) | 0.958(96) | 0.8982(61) | 0.2(10) |

Table 6: Fit results to Eq. (29). $\chi^2$ is computed per degree of freedom.

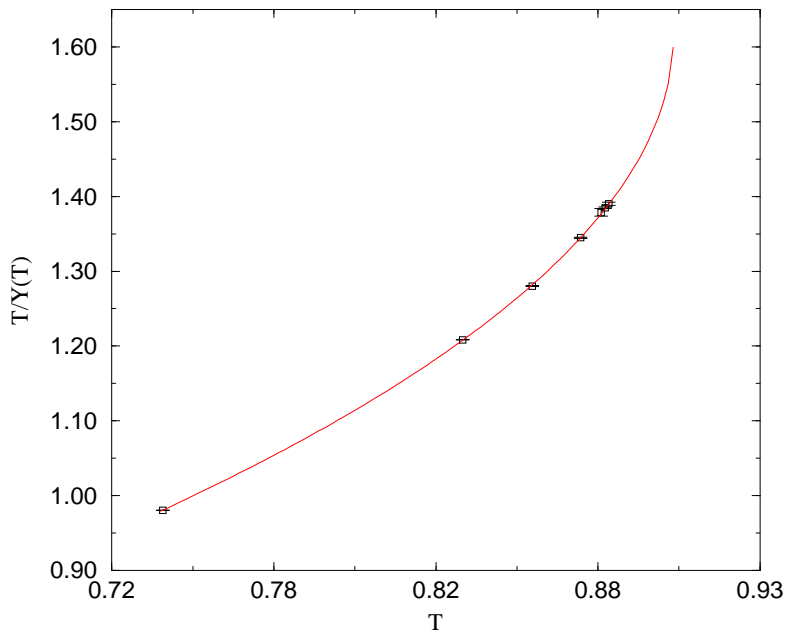

Figure 5: $T/\Upsilon(T)$ at $L = \infty$ as a function of $T$. The solid curve is the fit to (29).

| data points | $b$ | $T_c$ | $\chi^2$ |
|---|---|---|---|
| 7 | 0.91736(15) | 0.89016(4) | 63.13(63) |
| 6 | 0.8999(10) | 0.89220(13) | 0.01(71) |
| 5 | 0.9000(32) | 0.89219(24) | 0.02(82) |

Table 7: Fit results to Eq. (29) by taking $K'_c = \pi/2$. $\chi^2$ is computed per degree of freedom.

We would like to add a final comment. We have shown that because of the expression (10) for the temperature $T^*$ above $T_c$, where the KT correlation length becomes of the size of the system, significant finite size efffects such as those shown in Fig.1 appear in the measurement of the superfluid density. In fact, because of this very weak dependence of $T^* - T_c$ on $L$, finite-size effects in real 2D helium films should be observable for very large planar films. For example, taking $L$ as large as $100 cm$ we find that $T^*/T_c - 1 \sim 10^{-2} - 10^{-3}$ and this seems larger than the experimental resolution[12, 22].

# 6  Acknowledgements

This work was supported by the National Aeronautics and Space Administration under grant no. NAGW-3326.